\documentclass[twocolumn, prl, superscriptaddress]{revtex4-1}
\usepackage{amssymb}
\usepackage{amsmath}
\usepackage{epsfig}
\usepackage{color}
\usepackage{graphics, graphicx}
\usepackage{bbold}
\usepackage{psfrag}
\usepackage{mathcomp}
\usepackage{subfigure}
\usepackage{verbatim}
\usepackage{color}
\usepackage[colorlinks,citecolor=blue]{hyperref}

\begin{document}
\title{High order exceptional points in ultracold Bose gases }
\author{Lei Pan}
\affiliation{Beijing National Laboratory for Condensed Matter Physics, Institute of Physics, Chinese Academy of Sciences, Beijing 100190, China}
\affiliation{School of Physical Sciences, University of Chinese Academy of Sciences, Beijing 100049, China}
\author{Shu Chen}
\email{schen@iphy.ac.cn}
\affiliation{Beijing National Laboratory for Condensed Matter Physics, Institute of Physics, Chinese Academy of Sciences, Beijing 100190, China}
\affiliation{School of Physical Sciences, University of Chinese Academy of Sciences, Beijing 100049, China}
\affiliation{The Yangtze River Delta Physics Research Center, Liyang, Jiangsu 213300, China}
\author{Xiaoling Cui}
\email{xlcui@iphy.ac.cn}
\affiliation{Beijing National Laboratory for Condensed Matter Physics, Institute of Physics, Chinese Academy of Sciences, Beijing 100190, China}
\affiliation{Songshan Lake Materials Laboratory , Dongguan, Guangdong 523808, China}
\date{\today}

\begin{abstract}
We show that arbitrarily high-order exceptional points (EPs) can be achieved in a repulsively interacting two-species Bose gas in one dimension. By exactly solving the non-Hermitian two-boson problem, we demonstrate the existence of third-order EPs when the system is driven across the parity-time symmetry breaking transition. We further address the fourth-order EPs with three bosons and generalize the results to $N$-body system, where the EP order can be as high as $N+1$.
Physically, such high order originates from the intrinsic ferromagnetic correlation in spinor bosons, which renders the entire system collectively behave as a single huge spin.
Moreover, we show how to create ultra-sensitive spectral response around EPs via an interaction anisotropy in different spin channels.
Our work puts forward the possibility of atomic sensors made from highly controllable ultracold gases.
\end{abstract}

\maketitle

{\it Introduction.}
One of the most remarkable feature of non-Hermitian systems, as compared to Hermitian ones,  is their extreme sensitivity to external perturbations around the spectral degeneracy, which is known as the exceptional point (EP)\cite{Heiss1,Moiseyev,Kato}. For conventional degeneracy in Hermitian systems, any perturbation will produce an energy shift that at most linearly depends on the perturbation strength $\sim \epsilon$, and the shift becomes negligibly small $\sim \epsilon^n$ for high perturbation order $n$. While around an EP of $n$-th order, where $n$ is the number of energy levels that simultaneously coalesce, the perturbation can give rise to an energy shift $\sim \epsilon^{1/n}$, which grows as increasing $n$ and becomes greatly magnified for large $n$. Such sensitive response to tiny perturbations makes the non-Hermitian EP system an ideal candidate for sensors\cite{Wiersig1,Fleury,Wiersig2,Liu,Ding,Yang}. In the past few years, second-order EP ($n=2$) has been observed in various photonic, acoustic and atomic systems\cite{Dembowski,Dietz,Lee,Choi,Guo,Lin1,Feng1,Zhen,Sun,Doppler,Xu,Dembowski2,Yang2,Miao,Gao,
Hodaei1,Feng2,Ruter,Regensburger,Liertzer,Zhu,Brandstetter}. While higher-order EPs have been studied by a number of theoretical works\cite{Graefe, Demange,Teimourpour,Heiss2,Heiss3,Lin,Jing,Zhong}, their realizations in laboratories appear to be rather difficult. Very recently, two groundbreaking experiments have successfully achieved the third-order EPs and detected the enhanced sensitivity in coupled acoustic cavities\cite{Ding2} and optical micro-ring system\cite{Hodaei2}. Given the power-law growing sensitivity of EP sensors in terms of the associated EP order, the search for non-Hermitian systems with high-order EPs is strongly demanded.

In this work, we show how to achieve arbitrarily high-order EPs in an ultracold gas of spinor bosons. Specifically, we consider a two-species Bose gas in one dimension across the parity-time-reversal(PT) symmetry breaking transition, which can be experimentally realized by using an rf field in combination with laser-induced dissipations\cite{Luo}. We show that in the presence of spin-independent interactions, the EP order can be as high as $N+1$ with $N$ the total number of bosons. Such high order originates from the intrinsic ferromagnetic correlation in spinor bosons, which makes the entire many-body system collectively behave as a single huge spin. At these high-order EPs, the large energy degeneracy can be lifted up by fine-tuning the few-body coupling strength to be anisotropic in spin channels, which can be utilized for atomic sensors.
To demonstrate these results, we start with elaborating on the third-order EP by exactly solving the non-Hermitian two-boson problem, and then address the fourth-order EP with three bosons and finally approach to the many-body system.

{\it Two-body problem.} We consider two bosons in trapped 1D system with Hamiltonian $H=\sum_{i=1,2} H^{(0)}_i+U_{2b}$, ($\hbar=1$ throughout the paper)
\begin{eqnarray}
H^{(0)}_i&=&\sum_{\sigma} \left(-\frac{1}{2m}\frac{\partial^2}{\partial x_{i\sigma}^2}+\frac{1}{2}m\omega^2x_{i\sigma}^2\right)+H^{PT}_{i}; \nonumber\\
U_{2b}&=&\frac{1}{2}\sum_{i\neq j} \sum_{\sigma\sigma'}g_{\sigma\sigma'} \delta(x_{i\sigma}-x_{j\sigma'}).
\end{eqnarray}
Here $x_{i\sigma}$ is the coordinate of $i$-th particle with spin-index $\sigma=\uparrow,\downarrow$; $\omega$ is the harmonic frequency; $g_{\sigma\sigma'}$ is the coupling strength between spin $\sigma$ and $\sigma'$; the PT-symmetric potential is written as \cite{Luo}
\begin{equation}
H^{PT}_{i}=\Omega(s_{x,i}+i\Gamma s_{z,i}), \label{h_pt}
\end{equation}
with $s_{x,y,z}$ the spin-half operactors. In the single particle sector, a second-order EP occurs at $\Gamma=1$ where the two energy levels coalesce and the eigenstates undergo the PT-symmetry breaking transition\cite{Bender}. 

According to the Lippman-Schwinger equation, the two-body wave function $|\Psi\rangle$ satisfies
\begin{equation}
|\Psi\rangle=G_EU_{2b}|\Psi\rangle, \label{psi}
\end{equation}
where $G_E=(E-H^{(0)}_1-H^{(0)}_2)^{-1}$ is the non-interacting Green function, and $E$ is the eigen-energy. Since the  center-of-mass motion of two particles can be factored out, we only concentrate on their relative motion and the spin sector. By noting that $U_{2b}$ only acts on the spin-triplet space, we denote the relevant spin states as $|1\rangle\equiv |\uparrow\uparrow\rangle$, $|0\rangle\equiv (|\uparrow\downarrow\rangle+|\downarrow\uparrow\rangle)/\sqrt{2}$ and $|-1\rangle\equiv |\downarrow\downarrow\rangle$. Accordingly, $g_{\uparrow\uparrow},\ g_{\uparrow\downarrow},\ g_{\downarrow\downarrow}$  can be replaced by $g_1,\ g_0, \ g_{-1}$, respectively, denoting the  coupling strengths in $m=1,0,-1$ spin-triplet channels. Now we introduce three variables $\{f_{m}\}$ in 
\begin{equation}
\langle x|U_{2b}|\Psi\rangle=\sum_m f_m|m\rangle \delta(x), \label{U}
\end{equation}
with $x$ the relative coordinate of two bosons. Combining (\ref{psi}) and (\ref{U}), we arrive at three coupled equations in terms of $\{ f_m\}$, which gives the $E$-solution by solving:
\begin{equation}
{\rm Det}\left( \frac{1}{g_m}\delta_{mm'} -\langle m| G_E(0,0) |m'\rangle \right) =0.  \label{E}
\end{equation}
Here the Green function can be expanded as
\begin{equation}
G_E(x,x')=\sum_n \sum_j \frac{\psi_n(x)\psi^*_n(x')}{E_{rel}-E_n-\epsilon_j} \frac{|\mu_j^R\rangle \langle \mu_j^L|}{\langle \mu_j^L|\mu_j^R\rangle},  \label{G}
\end{equation}
where $E_{rel}=E-\omega/2$; $\psi_n(x)$ is the eigen-wavefunction for the relative motion with eigen-energy $E_n=(n+1/2)\omega$; $|\mu_j^R\rangle$ and $|\mu_j^L\rangle$ are the left and right spin vectors defined through $H_{PT}|\mu_j^R\rangle=\epsilon_j|\mu_j^R\rangle$ and $H_{PT}^\dagger|\mu_j^L\rangle=\epsilon_j^*|\mu_j^L\rangle$\cite{footnote}, here $H_{PT}=\sum_i H^{PT}_i$.
Note that the spin expansion in (\ref{G}) fails at the location of EP ($\Gamma=1$), where the single eigen-vector is inadequate to expand the whole spin space. Because of this, we have further carried out the exact diagonalization to solve the spectrum at $\Gamma=1$, and also confirmed that the two methods give consistent results in the regime $\Gamma\neq 1$.

\begin{figure}[t]
\includegraphics[width=8.5cm]{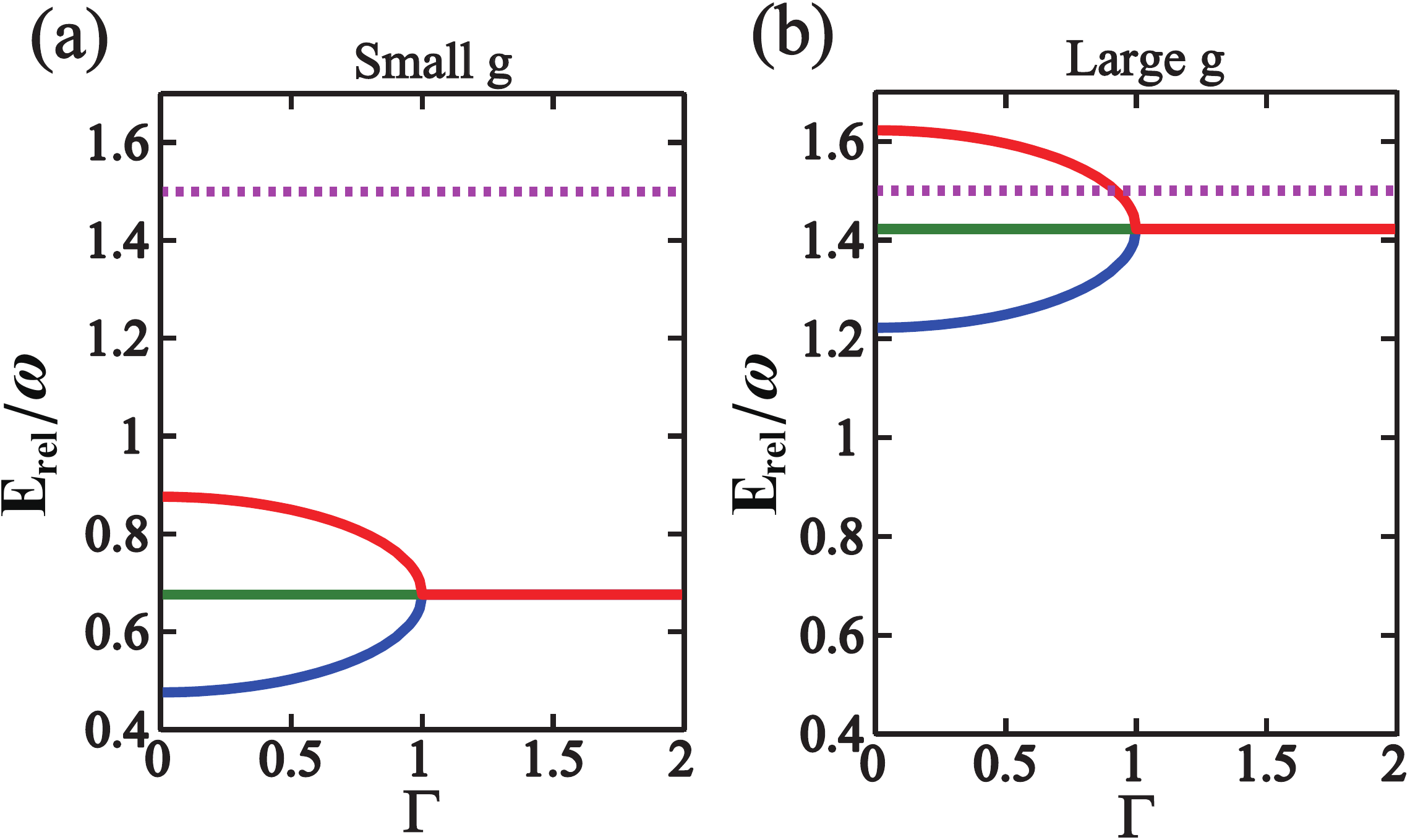}
\caption{(Color online). Exact solution of the lowest four energy levels for two bosons with isotropic interactions $g_1=g_0=g_{-1}\equiv g$. (a) is for weak coupling $g=0.5\omega l$, (b) is for strong coupling $g=20\omega l$. The horizontal dashed lines in (a) and (b) are for the spin-singlet state, which is immune to interactions. A third-order EP appears in both (a) and (b) at $\Gamma=1$. Here $l=1/\sqrt{m\omega}$ is the confinement length and we take  $\Omega=0.2\omega$.} \label{fig_isotropic}
\end{figure}

In Fig.\ref{fig_isotropic}, we plot the lowest four energy levels for isotropic interactions, $g_1=g_0=g_{-1}\equiv g$, in both weak (a) and strong (b) coupling regime. We see that in both (a) and (b), the lowest three energy levels merges at $\Gamma=1$, beyond which the upper and lower energies start to develop imaginary parts, and meanwhile, all the three eigenvectors also coalesce at $\Gamma=1$. These are all characteristic features of a third-order EP. Such third order can be further checked through the spectral response to small perturbations, as shown below.

\begin{figure}[t]
\includegraphics[width=8.5cm]{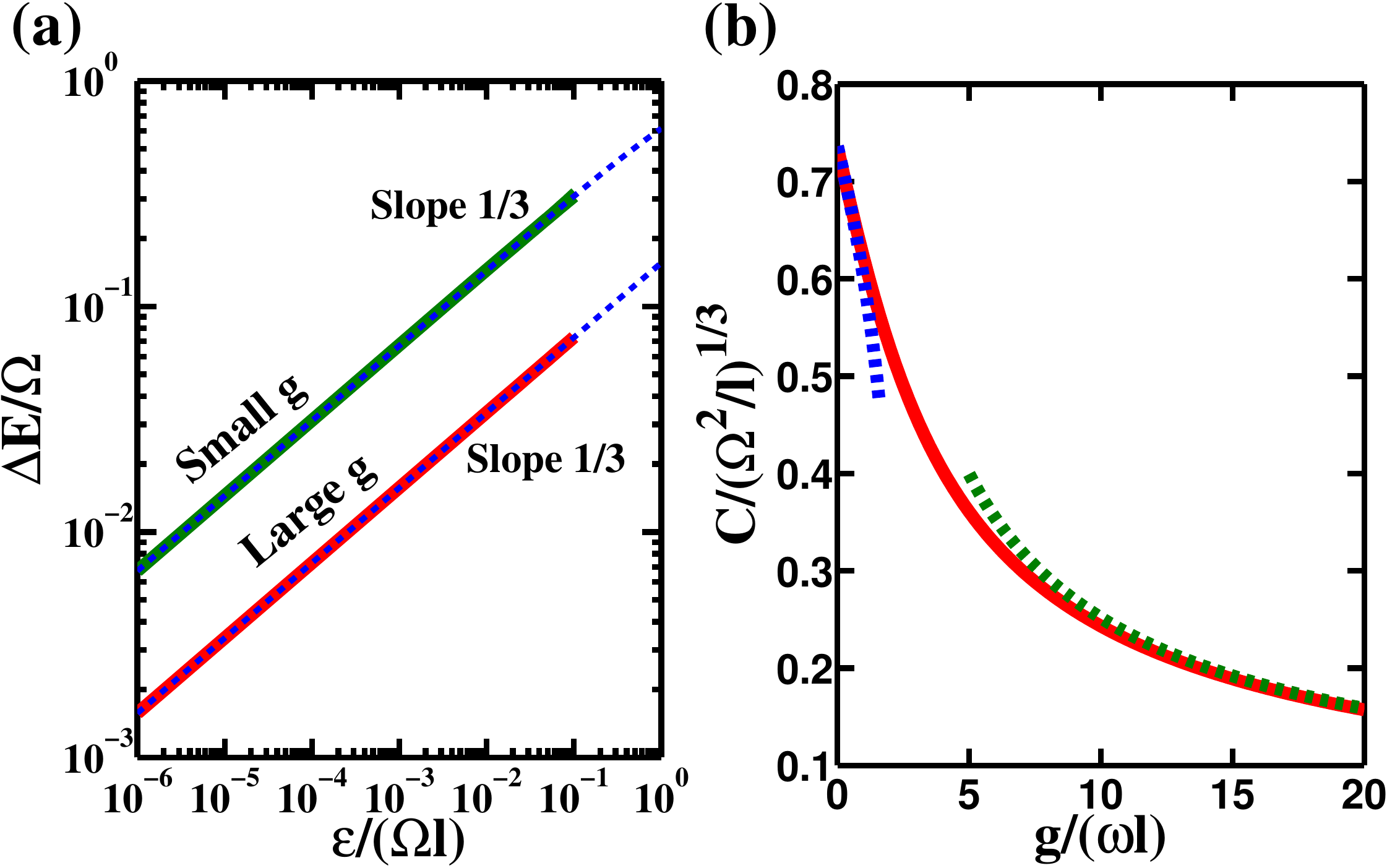}
\caption{(Color online). (a) Energy splitting ($\Delta E$) of two bosons at $\Gamma=1$ as the function of interaction isotropy ($\epsilon$) in $m=0$ channel.  Upper green and lower red solid lines  are for weak coupling $g=0.5\omega l$ and strong coupling $g=20\omega l$.  Blue dashed lines are fitting curves according to the cube-root relation $\Delta E= C \epsilon^{1/3}$ (Eq.\ref{cuberoot}). (b) The coefficient $C$ as function of $g$. Blue and green dashed lines are respectively obtained from second-order perturbation theory in small $g$ and effective spin chain model in large $g$ limits (see text). Here $\Omega=0.2\omega$.} \label{fig_anisotropic}
\end{figure}

We introduce external perturbations through the interaction anisotropy in spin channels, which is easy to implement in cold atoms by tuning magnetic field. Here we take, for instance, a tiny interaction anisotropy in $m=0$ channel, i.e., $g_{1}=g_{-1}=g$ and $g_{0}=g+\epsilon$. The exact solution shows that the original degenerate energy levels at $\Gamma=1$ split 
with the same amplitude $|\Delta E_{i=1,2,3}|\equiv \Delta E$. In Fig.\ref{fig_anisotropic}(a), we plot $\Delta E$ as a function of $\epsilon$, where a cube-root relation can be identified in all coupling regime:
\begin{equation}
\Delta E= C \epsilon^{1/3}. \label{cuberoot}
\end{equation}
This relation ultimately confirms the existence of third-order EP in two-boson system. In Fig.\ref{fig_anisotropic}(b), we further plot the coefficient $C$ as a function of $g$. The asymptotic behaviors of $C$ in weak and strong $g$ limits will be discussed later. As a comparison, we note that the ground state of the Hermitian system is three-fold degenerate when $\Omega=0$ and $\epsilon=0$, and the introduction of anisotropic interaction would split the triple state with induced energy splitting $\Delta E \propto \epsilon$. This suggests that the energy splitting around the third-order EP is much more sensitive to the tiny anisotropic interaction than the corresponding Hermitian system, which can be exploited for ultra-sensitive sensing.

A remarkable result shown above is that, given the non-Hermitian potential (\ref{h_pt}), the order of EP at $\Gamma=1$ can be upgraded from $2$ to $3$ when the boson number increases from $1$ to $2$. Physically, this order-upgrading can be traced back to the intrinsic ferromagnetic correlation in spin-1/2 bosons\cite{Li, Guan}. It can be seen easily in the strong coupling regime, where the system can be described by an effective ferromagnetic spin chain $H=-J\sum_{\langle i,j\rangle}{\bf s}_i\cdot {\bf s}_j$ ($J>0$)\cite{Furusaki, Cui}, resulting in a ferromagnetic ground state.
Since the PT potential $H_{PT}$ commutes with the total spin, the ferromagnetic  state is also the eigen-state of $H_{PT}$. In the case of two bosons, the ferromagnetic state is spin-triplet ($S=1$) with three components, and in this subspace  
the operators $S_{\alpha}=\sum_i s_{\alpha,i}$ in $H_{PT}$ just behave as spin-1 operators. Equivalently, the two bosons constitute a spin-1 object, and accordingly the EP order is  upgraded to $2S+1=3$.

Given above picture, the energy splitting under a small interaction anisotropy (see Eq.\ref{cuberoot}) can be analyzed by expanding the two-body Hamiltonian only in spin-triplet space. In weak coupling limit, a second-order perturbation theory  based on unperturbed non-interacting system gives the cube-root relation (\ref{cuberoot}) with $C=\big[\frac{\Omega^2}{\sqrt{2\pi}l}(1-\frac{\sqrt{2}g\gamma}{\sqrt{\pi}\omega l})\big]^{1/3}$ where $\gamma\approx0.577$ is the Euler constant. In strong coupling limit, we resort to the effective spin-chain model for spin-1/2 bosons \cite{Cui}:
\begin{equation}
H_{\rm eff}=J\left( -\frac{1}{g} {\bf s}_1\cdot {\bf s}_2- \frac{2\epsilon}{g^2} s_{z,1}s_{z,2} \right) +\sum_i H^{\rm PT}_i. \label{H_eff}
\end{equation}
Here we have assumed $1/g_1-1/g\sim -\epsilon/g^2$. Expanding (\ref{H_eff})  in spin-triplet states, we obtain $\Delta E$ following Eq.\ref{cuberoot} with $C=\big[\frac{J\Omega^2}{g^2}\big]^{1/3}$. These asymptotic behaviors of $C$ in weak and strong $g$ limits can well fit the exact results, see Fig.\ref{fig_anisotropic}(b). In addition, we have tried interaction anisotropies in other spin channels ($m=1,-1$), and found the cube-root relation and the asymptotic behaviors of $C$ are not qualitatively altered.

{\it Three-body system.} We now turn to three-boson problem. In the presence of an spin-independent interaction, it is easily drawn from previous analysis that the ground state is ferromagnetic with total spin $S=3/2$, and the PT potential  will result in an EP at $\Gamma=1$ with order $2S+1=4$. It is then promising to achieve an even sensitive spectral response as $\Delta E\sim \epsilon^{1/4}$, given that a proper perturbation is introduced. In the following, we will show that such a fourth-root sensitivity can be induced by an anisotropy in three-body couplings.

We consider three trapped bosons experiencing small interaction anisotropy in, for instance, two-body $\uparrow\downarrow$ and/or three-body $\uparrow\downarrow\downarrow$ scattering channels. To simplify the analysis while keeping the essence of physics, we concentrate on the strongly repulsive regime (with large two-body repulsion in all channels), where the system can be described by following effective spin chain:
\begin{eqnarray}
H_{\rm eff}&=&\sum_{i=1}^2 \left(-\frac{J}{g} {\bf s}_i\cdot {\bf s}_{i+1} + \epsilon_2  s_{z,i}s_{z,i+1}\right) + \epsilon_3 s_{z,1}s_{z,2}s_{z,3}\nonumber\\
 && + \sum_{i=1}^3 H^{\rm PT}_i. \label{H3}
\end{eqnarray}
Here 
$\epsilon_2$ and $\epsilon_3$ respectively refer to the two-body and three-body interaction anisotropies.  In writing (\ref{H3}), we have omitted the term $\sim \epsilon_3\sum_i s_{z,i}$, as it does not contribute to the sensitive spectral response and can be eliminated by an additional tiny magnetic field.

\begin{figure}[t]
\includegraphics[width=8.5cm]{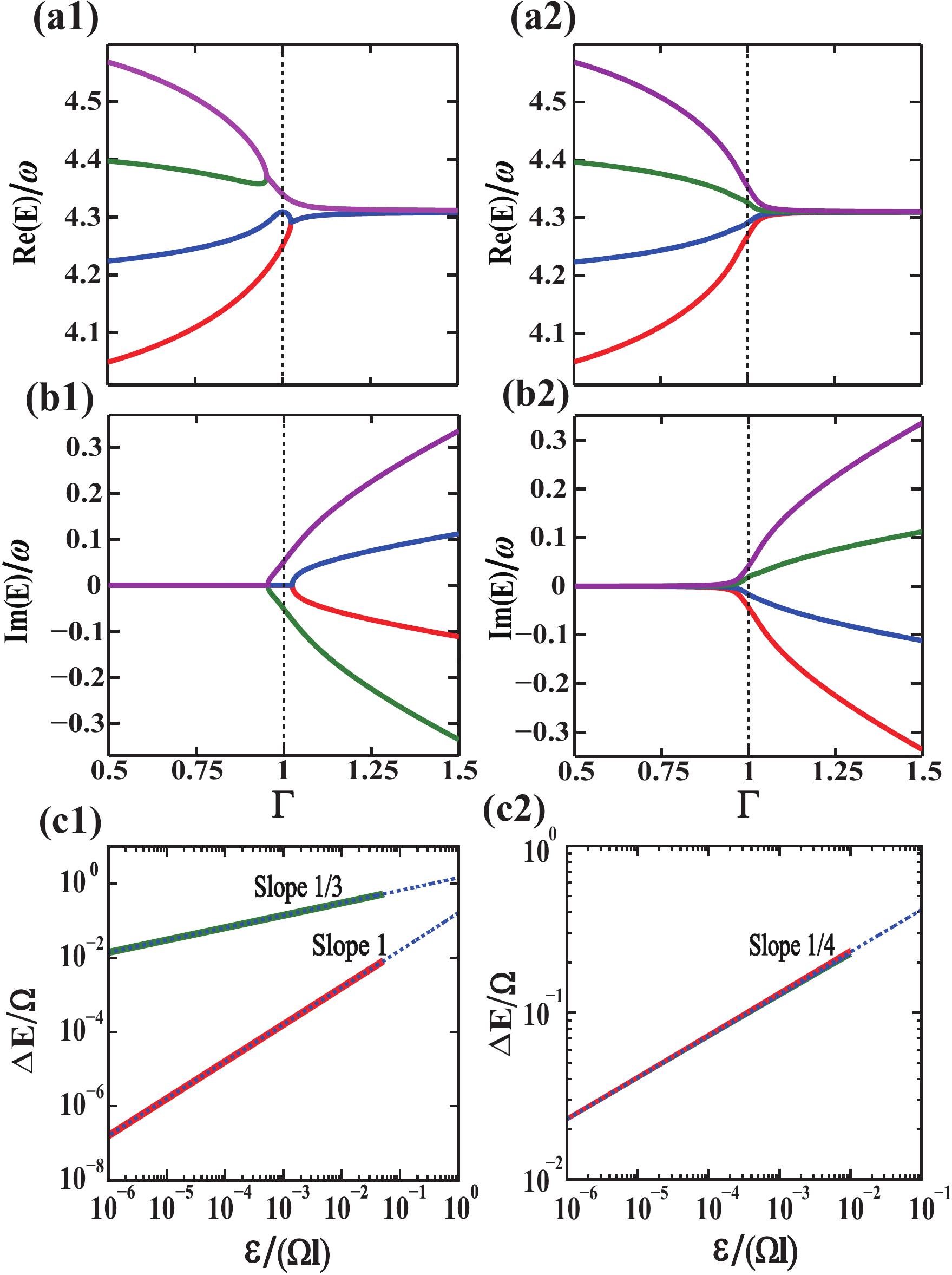}
\caption{(Color online). Spectral response for the lowest four energy levels in three-boson system to different types of interaction anisotropies. The real and imaginary parts of the energies are shown as a function of $\Gamma$ only with two-body anisotropy $\epsilon_2/(\Omega l)=0.01,\epsilon_3/(\Omega l)=0$ (a1,b1) where we can find a pair of complex conjugate eigenvalues with the same real parts and two purely real eigenvalues on the line $\Gamma=1$, or only with  three-body anisotropy $\epsilon_3/(\Omega l)=0.01, \epsilon_2/(\Omega l)=0$ (a2,b2) where two pairs of complex conjugate eigenvalues appear simultaneously. Accordingly, the energy shift at $\Gamma=1$ is plotted as a function of $\epsilon_2$ (c1) or $\epsilon_3$ (c2). Here $\Omega=0.2\omega$.} \label{fig_3body}
\end{figure}

In Fig.\ref{fig_3body}, we show the spectral response for the lowest four energy levels to different types of interaction anisotropies. Depending on the anisotropy from two-body ($ \epsilon_2\neq 0$, $\epsilon_3=0$) or from three-body ($\epsilon_3\neq 0$, $\epsilon_2=0$) sector, the spectral response shows distinct structures around  $\Gamma=1$. In the case of only $\epsilon_2\neq 0$, at $\Gamma=1$ three different values are left for the real and imaginary parts of the energies, see Fig.\ref{fig_3body}(a1),(b1); accordingly, the original fourth-order EP splits to a third-order one with cube-root dependence and a trivial one with linear dependence, see (c1). In the case of $\epsilon_3\neq 0$, the real and imaginary parts of four energies all split at $\Gamma=1$ (see (a2)(b2)), and the fourth-root scaling can be achieved (see (c2)). That is to say, to optimize the spectrum sensitivity near the fourth-order EP, i.e., to realize $\Delta E\sim \epsilon^{1/4}$,  a three-body interaction anisotropy is a crucial ingredient.

{\it Many-body system.} Now we generalize above discussions to two-species boson system with arbitrary particle number $N$ and under $M$-body interactions.

\begin{figure}[t]
\includegraphics[width=8.5cm]{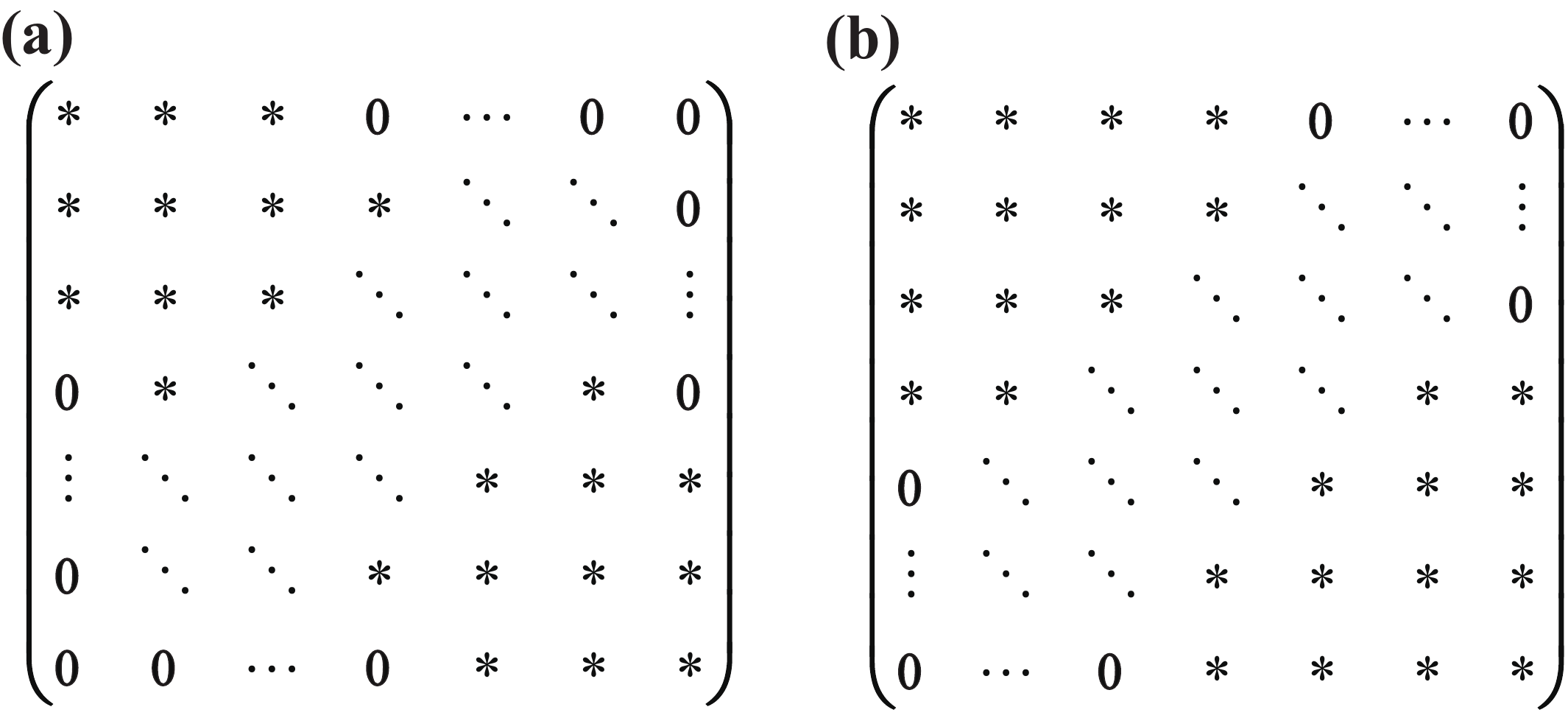}
\caption{(Color online). General $(N+1)\times(N+1)$ Hamiltonian matrix  for N-bosons in the (rotated) ferromagnetic spin space. Here $"\ast"$ denotes non-zero element. (a) and (b) are respectively with two-body and three-body interaction anisotropy. } \label{fig_matrix}
\end{figure}

First, in the presence of spin-independent interaction which supports a ferromagnetic ground state, the system collectively behaves as a single huge spin with $S=N/2$ and a high EP order $2S+1=N+1$ can be achieved. For the convenience of later discussion, we introduce an alternative way to understand this result. At $\Gamma=1$, we have $H_{PT}=\Omega (S_x+iS_z)$, with $S_{\alpha}=\sum_i s_{i,\alpha}$ the spin-$N/2$ operators. Under a spin rotation around $x$, $H_{PT}$ simply reproduces the angular momentum raising operator $S_+=S_x+iS_y$. Such operator can be expanded in $\{S_z\}$ space as a $(N+1)\times(N+1)$ matrix, which has one single eigenvalue ($=0$) and one single eigen-vector ($|S_z=N/2\rangle=|1,0,...0\rangle$). This justifies the occurrence of $(N+1)$-th order EP in $(N+1)$-dimensional spin space.


Secondly, when turn on a small $M$-body interaction isotropy, the original $(N+1)$-th order EP will generally split into a number of sub-EP groups depending on the values of $M$ and $N$. To see this, again we resort to the effective model in the strong coupling regime and work only within $S=N/2$ subspace, where the spin-dependent Hamiltonian at EP can be generally written as
\begin{equation}
H_{sd}= \Omega (S_x+iS_z)+ \epsilon \sum_i c_i \prod_{j=0}^{M-1} s_{z, i+j} . \label{h_many}
\end{equation}
Here $c_i$ is the position-dependent coupling constant  due to the trapping potential, and we have omitted other less important terms $\sim \prod_{j=0}^{n} s_{z, i+j}$ with $n<M-1$, which produces less sensitive spectral response. Again under a spin rotation around $x$, the PT term becomes $S_+$ operator, and the perturbation terms become $M$-rank polynomials in terms of $s_{y,i}$, which in the ferromagnetic subspace will generate terms like $S_{\pm}^m$ (with $m\le M$). In Fig.\ref{fig_matrix} (a) and (b), we show the typical structures of Hamiltonian matrix for $M=2$ and $M=3$, where the non-zero elements can at most extend to the second(for $M=2$)  or the third(for $M=3$) super- and sub-diagonals. According to a mathematic study in Ref.\cite{Ma}, this is the structure of Jordan blocks $J_{N+1}$ with perturbations constituting the $(M+1)$-Hessenberg matrix,
under which the $(N+1)$-th order EP splits to $[\frac{N+1}{M+1}]$ groups of sub-EP and each with order $M+1$. That is to say,  a tiny perturbation $\epsilon$ in the $M$-body couplings can generate an energy splitting as $\epsilon^{1/(M+1)}$ in the eigen-spectrum of $N$-boson system ($N\ge M$). This covers our previous analyses on the spectral response to  the two- and three-body interaction anisotropies. 

{\it Experimental relevance.} Experimentally, a two-species Bose gas with nearly spin-independent interaction can be achieved by using the lowest two hyperfine states of $^{87}$Rb atoms, i.e., $|\hspace{-0.15cm}\uparrow\rangle=|F=1, m_F=0\rangle$ and $|\hspace{-0.15cm}\downarrow\rangle=|F=1, m_F=-1\rangle$, where the bare scattering lengths in different spin channels are rather close\cite{Rb87}. The two-body interaction anisotropy can be further fine-tuned through the magnetic field. By applying a rf field to couple these two states and tune the rf frequency to match their Zeeman splitting, the transverse ($\sigma_x$) field can be realized, and meanwhile, the third hyperfine state $|F=1, m_F=-1\rangle$ can be adiabatically eliminated due to the finite quadratic Zeeman energy. The non-Hermitian term ($i\sigma_z$) can be implemented by laser-induced dissipations\cite{Luo}.
To generate the three-body interaction, one can tune the magnetic field nearby an Efimov resonance in particular collision channel\cite{Braaten, Grimm}, or directly utilize the transverse confinement to create visible three-body strengths in quasi-1D geometry\cite{Mazets, Pricoupenko, Nishida, Petrov}. The spectral response discussed in this work can be easily measured in cold atoms experiment using rf spectroscopy.

{\it Summary.} In summary, we have demonstrated the existence of arbitrarily high order EPs in the non-Hermitian 1D two-species Bose gas. This is facilitated by the ferromagnetic correlation in interacting spinor bosons, such that the EP order directly scales as the number of bosonic atoms. The scheme is thus substantially easier to implement as compared to previous ones in other systems creating high-order EPs. Moreover, we have pointed out that  a small interaction anisotropy in spin channels can be used to generate ultra-sensitive spectral response. Specifically, a two-body (three-body) interaction anisotropy is responsible for a cube-root (fourth-root) spectral response. Our work thus can serve as a guideline for making sensors based on ultracold atoms. Stimulated by this work, in future it is interesting to explore more intriguing physics due to the interplay of non-Hermitian potentials and strong interactions.

{\bf Acknowledgement.} The work is supported by the National Key Research and Development Program of China (2018YFA0307600, 2016YFA0300603), and the National Natural Science Foundation of China (No.11622436, No.11425419, No.11421092, No.11534014).


\begin{thebibliography}{99}
\bibitem{footnote} By this definition, one can prove $\sum_j |\mu_j^R\rangle \langle \mu_j^L|/\langle \mu_j^L|\mu_j^R\rangle$ is an identity matrix expanded in spin space. This is true except at EP (when $\Gamma=1$).
\end{thebibliography}
\end{document}